\documentclass[aps,pre,twocolumn,superscriptaddress]{revtex4}

\usepackage{graphicx}

\usepackage{amsmath, amssymb} 

\usepackage{color}

\DeclareMathOperator{\Tr}{Tr}

\begin{document}

\title{Entanglement Spectrum and Entanglement Hamiltonian of a Chern insulator with open boundaries}

\author{Maria Hermanns}
\affiliation{Institut f\"ur Theoretische Physik,
Universit\"at zu K\"oln, Z\"ulpicher Stra\ss e 77, 50937 K\"oln, Germany}

\author{Yann Salimi}
\affiliation{Institut f\"ur Theoretische Physik,
Universit\"at zu K\"oln, Z\"ulpicher Stra\ss e 77, 50937 K\"oln, Germany}

\author{Masudul Haque}
\affiliation{Max Planck Institute for the Physics of Complex Systems, N\"othnitzer Str. 38, 01187
  Dresden, Germany}

\author{Lars Fritz}
\affiliation{Institut f\"ur Theoretische Physik,
Universit\"at zu K\"oln, Z\"ulpicher Stra\ss e 77, 50937 K\"oln, Germany}
\affiliation{Institute for Theoretical Physics and Center for Extreme Matter and Emergent Phenomena,
Utrecht University, Leuvenlaan 4, 3584 CE Utrecht, The Netherlands}

\date{\today}

\begin{abstract}

  We study the entanglement spectrum of a Chern insulator on a cylinder geometry, with the cut
  separating the two partitions taken parallel to the cylinder edge, at varying distances from the
  edge.  In contrast to similar studies on a torus, there is only one cut, and hence only one
  virtual edge mode in the entanglement spectrum.  The entanglement spectrum has a gap when the cut
  is close enough to the physical edge of the cylinder such that the edge mode spatially extends
  over the cut.  This effect is suppressed for parameter choices where the edge mode is sharply
  localized at the edge. In the extreme case of a perfectly localized edge mode, the entanglement
  spectrum is gapless even if the smaller partition consists of a single edge row.  For the
  single-row cut, we construct the corresponding entanglement Hamiltonian, which is a
  one-dimensional tight-binding Hamiltonian with complex long-range hopping and interesting properties.  
  We also study and explain the effect of two different schemes of flux insertion through a ring described by such an entanglement Hamiltonian.

\end{abstract}

\maketitle


\maketitle

\section{Introduction}

The study of insulating topological states of matter has emerged as an intensely active field of
research.  For non-interacting systems there is a variety of topological states which have been
identified and subsequently classified based on their elementary symmetries \cite{Classification1,Classification2}.
The most prominent example of this kind are the integer quantum Hall insulator states \cite{IQH}.
Many other symmetry protected topological insulator states, such as for instance quantum spin Hall
systems or three dimensional $Z_2$ topological insulators have been identified theoretically
\cite{Haldane1988, TopIns1,TopIns2,TopIns3} and also experimentally realized \cite{Exp1,Exp2}.

The \emph{entanglement spectrum} (ES) \cite{LiHaldane} has attracted substantial attention as a theoretical tool to
characterize and identify topological states.  Given a partition of the system into parts 1 and
2, the entanglement spectrum is (the negative logarithm of) the spectrum of the reduced density
matrix of part 1, 
which is obtained by tracing out the degrees of freedom of part 2 from the density matrix
corresponding to the system ground state.
Ref.~\cite{LiHaldane} introduced the notion that the ES for a topological state contains a
representation of the physical edge modes.  Even though the system may not possess an edge (e.g.,
due to periodic boundary conditions), the cut separating the partitions serves as a virtual edge for
the ES.  
This notion is important and relevant both for interacting topological states like
fractional quantum Hall states and spin liquids \cite{ES_in_FQH, ES_in_interacting_nonFQH} and for
non-interacting topological states \cite{EntSpecIntTI1, Fidkowski, HughesProdanBernevig,
  Alexandradinata@al, FangGilbertBernevig_PRB2013, TurnerZhangVishwanath_PRB2010,
  ProdanHughesBernevig_PRL2010, KargarianFiete_PRB2010, HuangArovas_PRB2012}.
The entanglement spectrum can be regarded as the spectrum of a `Hamiltonian' that has support on part 1.  
This entanglement Hamiltonian can be thought of as possessing some of the properties
of the physical Hamiltonian of the system, albeit with a boundary due to the restriction to part 1.  
When the system ground state is topological, the entanglement Hamiltonian is argued to also be a Hamiltonian with
topological properties, and since part 1 has a boundary, the ES can be expected to display an edge mode.  
The general properties of the entanglement Hamiltonian, and the physical content of the ES, are not
completely understood and remain a topic of active research.

In this work, we examine the entanglement spectrum of a two-dimensional Chern insulator on a
cylinder geometry, i.e., on a system having periodic boundary conditions in one spatial direction but
not in the other, similar to Ref. \cite{Alexandradinata@al}.  In such a system, there are
\emph{physical} chiral edge modes, in addition to the virtual edge modes present in the ES.  We
consider cuts parallel to the edges.  By varying the distance to the edge, this setup allows us to
study the interplay of the physical edge modes and the partitioning.  In particular, when the
smaller partition is small enough, the entanglement Hamiltonian should encode the physics of a
single (virtual) chiral edge mode.  
This is a rather unusual opportunity, because chiral edge modes
generally occur in pairs.  We demonstrate that, when the cut is near enough to the edge so that the
edge modes spatially cross the cut, the ES generically becomes gapped.  This is analogous to
topological matter on a thin strip, when the physical edge modes on opposite edges have the
opportunity to hybridize and produce a gap in the spectrum.  
We use a Chern insulator model for
which the localization of the boundary modes can be parametrically tuned, in particular, they can be
made to localize sharply at the physical edge.  At such special points, we find that the ES remains
gapless, containing a single chiral channel, even when the cut is such that the partition contains
only a single row at the edge.

Partitioning the system, such that the smaller subsystem contains only a single row, yields an
effectively one-dimensional system.  Since the ES has a chiral channel, the entanglement Hamiltonian
must correspond to a 1D chiral Hamiltonian, accomplished by complex long-ranged hopping.  In this
article, we study the properties of such entanglement Hamiltonians and argue that their ground
states carry a so-called `persistent current' \cite{persistentcurrent}. We also study the effect of
two different schemes of `flux insertion' through the cylinder on the entanglement spectrum.

\begin{figure}
\includegraphics[width=0.48\textwidth]{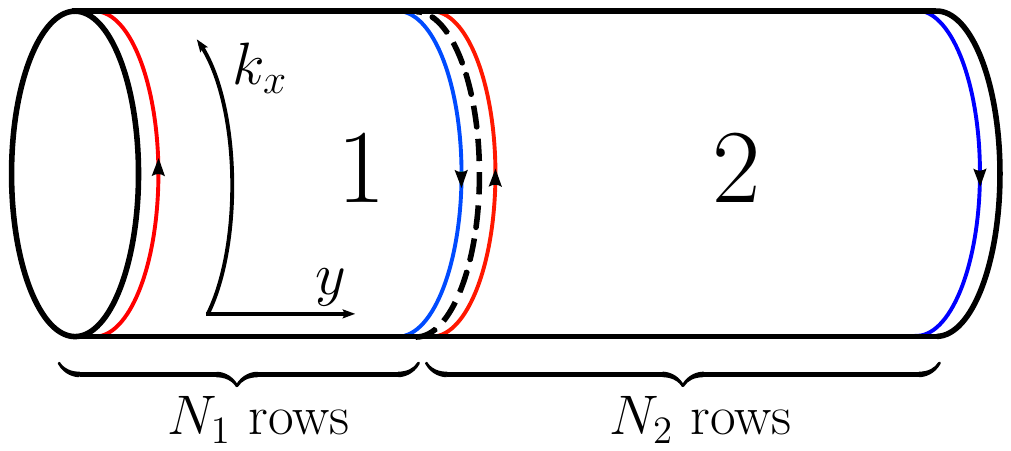}
\caption{Cylinder with partitions 1 and 2, between which the entanglement is calculated. The cut is
  parallel to the physical edges and is at variable distance from the edges; we choose $N_1<N_2$.
}\label{fig:cylinder}
\end{figure}

\paragraph*{Outline ---}
In Section \ref{sec:mandm}, we introduce the tunable model Hamiltonian we consider, and provide the
definitions we use for the entanglement spectrum and the entanglement Hamiltonian.  We use a
`single-particle' version of the ES, as is conventional in the literature on non-interacting
topological insulators. In the same spirit, we also introduce a single-particle version of the
entanglement Hamiltonian, which is a tight-binding hopping Hamiltonian.  In Section \ref{sec:ES}, we
present the ES of a a cylindrical Chern insulator as a function of the distance of the cut towards
an edge, see Fig.~\ref{fig:cylinder}.  We characterize the gap in the ES and relate it to the size
of the partition (distance between physical edge and entanglement cut) and the localization length
of the edge mode. In Section \ref{sec:entHam}, we extract and present the most prominent features of
the entanglement Hamiltonian in the extreme case where the smaller partition contains just one row.
Section \ref{sec:flux} introduces the two schemes of flux insertion on the cylinder, discusses their
properties and shows their relation to previous flux insertion and spectral flow studies on the
torus \cite{spectralflow,HughesProdanBernevig}.  In contrast to the torus we find that no spectral
flow is admitted on finite size systems.  However, in the thermodynamic limit, spectral flow can
occur for certain fine-tuned entanglement Hamiltonians.

\section{Model Properties and Entanglement Definitions}\label{sec:mandm}

\subsection{The two-orbital Chern insulator}

\begin{figure}
\includegraphics[width=0.45\textwidth]{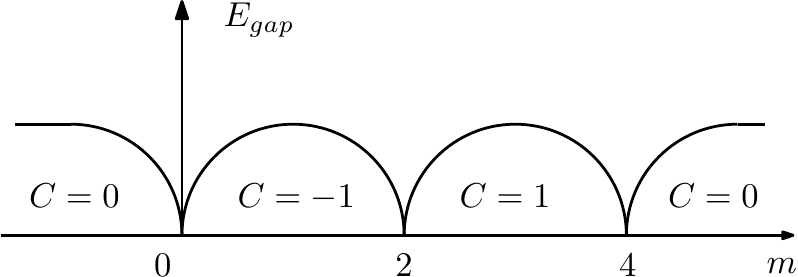}
\caption{Sketch of the phase diagram of the two-orbital Chern insulator model, Eq.~\eqref{eq:model}. The x-axis
  denotes the tuning parameter $m$. The phases are characterized by their respective Chern numbers
  $C$. Non-zero values of $C$ signal topological phases with gapless chiral edge
  modes.}\label{fig:pd}
\end{figure}

\begin{figure}
\includegraphics[width=0.5\textwidth]{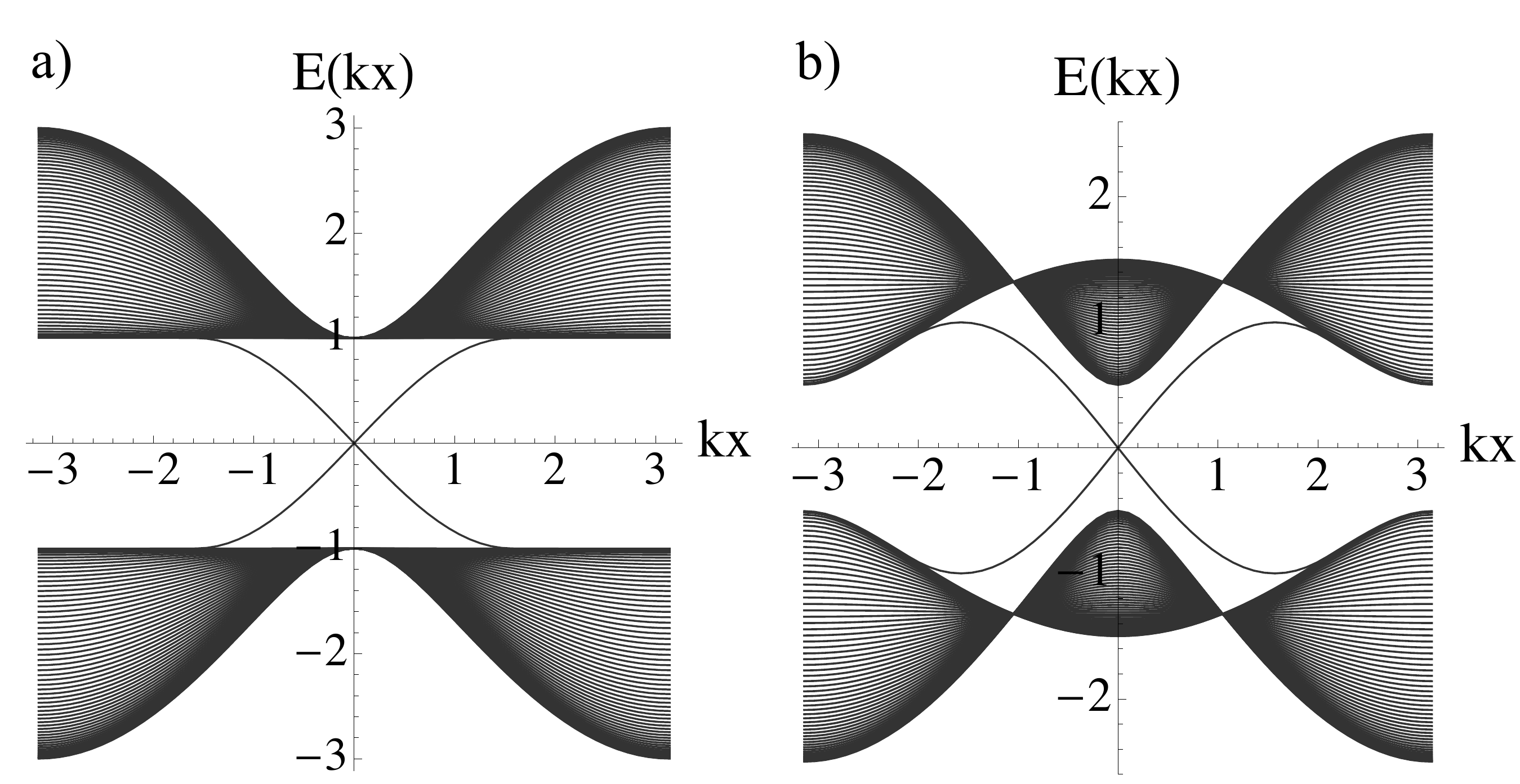}
\caption{Energy spectrum of the two-orbital Chern model on the cylinder geometry for a) $m=1$ and b)
  $m=1.5$. The x-axis denotes the momentum along the cylinder circumference.  The occurrence of the
  edge modes in the bulk gap is visible for both cases.  For the special point $m=1$ the edge modes
  are sharply localized due to the flat dispersion of the bulk bands.  The other case, $m=1.5$, can
  be regarded as a representative generic situation. }\label{fig:spectrumob}
\end{figure}

Underlying our study in this paper is the so-called two-orbital Chern insulator
\cite{HughesProdanBernevig}. The hopping Hamiltonian is given by
\begin{eqnarray}\label{eq:model}
\hat{H}&=&\sum_{\vec{R}}\left(\Psi^\dagger_{\vec{R}}\frac{i\sigma_x-\sigma_z}{2}\Psi^{\phantom{\dagger}}_{\vec{R}+x}+\Psi^\dagger_{\vec{R}}\frac{i\sigma_y-\sigma_z}{2}\Psi^{\phantom{\dagger}}_{\vec{R}+y}+\rm{h.c.}\right) \nonumber \\ &+&(2-m)\sum_{\vec{R}}\Psi^\dagger_{\vec{R}} \sigma_z \Psi^{\phantom{\dagger}}_{\vec{R}}\;.
\end{eqnarray}
where $\sigma_{x,y,z}$ are the standard Pauli matrices acting in the orbital space (the wave function $\Psi$ is a spinor whose entries refer to the two orbitals involved) and $m$ is a tuning parameter for the model (we assume henceforth a lattice constant $a=1$). The model has trivial and non-trivial insulating phases: For $0<m<2$
we have a band insulator with Chern number $-1$, for $2<m<4$ we have a band insulator with Chern
number $+1$ while for $m>4$ and $m<0$ we have a trivial band insulator with Chern number zero. At $m=0,2,4$ the bulk spectrum is gapless and the system exhibits semimetallic behavior. The phase
diagram is shown in Fig.~\ref{fig:pd} schematically showing the bulk gap as a function of $m$ and
the respective Chern numbers $n$.
In the regime $0<|m|<4$ (except for $m=2$ where the bulk itself is
gapless) we have chiral edge channels on the opposing edges of the cylinder.  This is shown in
Fig.~\ref{fig:spectrumob} which shows the spectrum of a system on a cylindrical geometry (compare
Fig.~\ref{fig:cylinder}) with mass parameter $m=1$ and $m=1.5$ in the Chern insulating phase.

\begin{figure}
\includegraphics[width=0.45\textwidth]{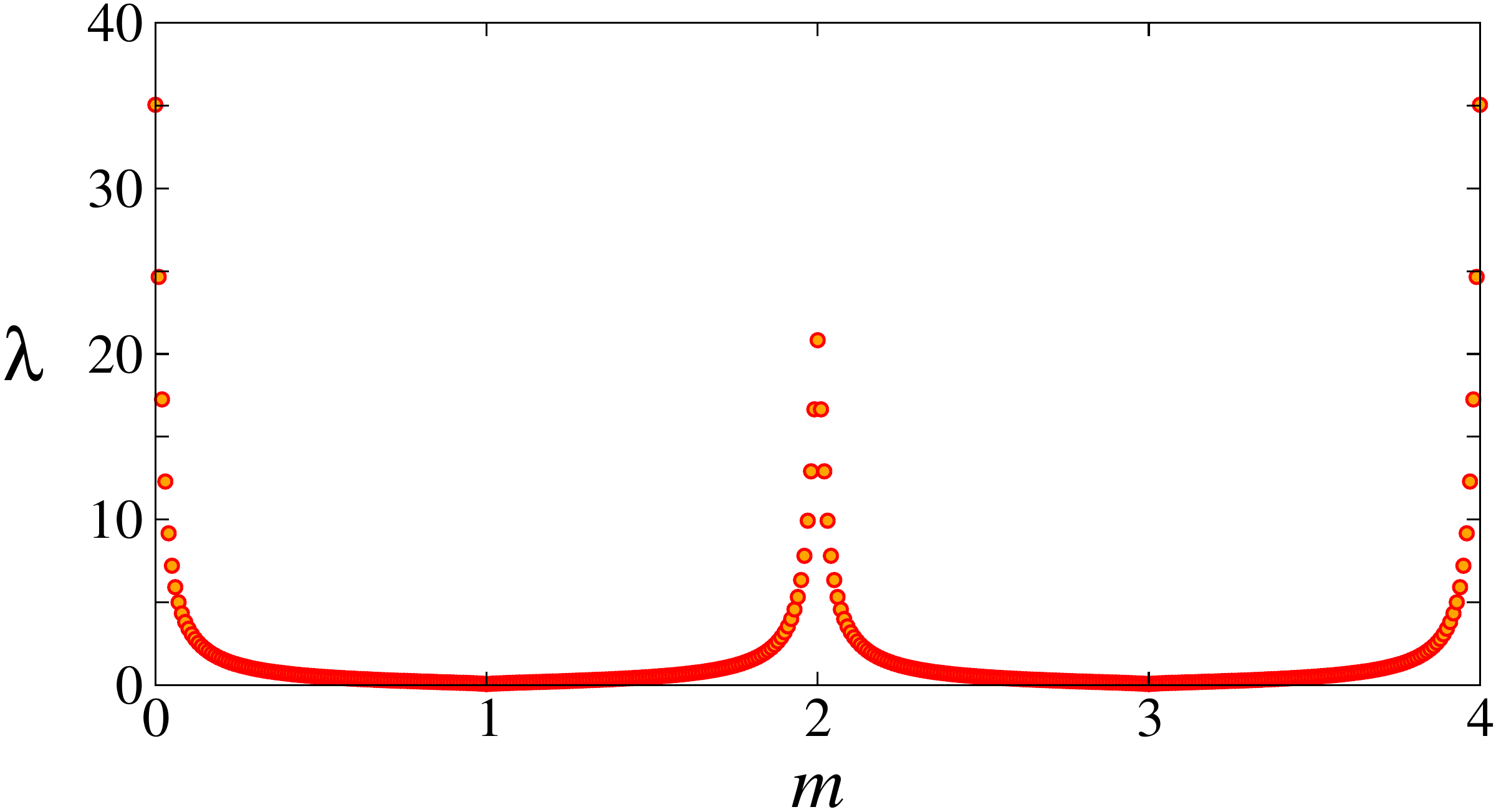}
\caption{ \label{fig:loclength_vs_m}
Localization length $\lambda$ of boundary states versus $m$.  The localization length
$\lambda=\ln2/\beta$ is obtained by fitting $\sim\exp(-\beta{y})$ to the single-particle eigenstate
just below the chemical potential, $|\psi(k'_x,y)|^2$. 
}
\end{figure}

The decay of the boundary modes into the bulk depends on the parameter $m$. When defining the corresponding localization length $\lambda$, we consider only the single-particle eigenstate directly below the chemical potential, as it turns out to be the boundary mode that is most relevant for the properties of the ES. 
As a measure for the localization length
we use the square of the wavefunction and determine the distance $\lambda$ when it has decayed to half
its original value.  Fig.~\ref{fig:loclength_vs_m} shows the behavior of $\lambda$ as a
function of the mass parameter $m$. We note that $\lambda$ decreases upon approaching $m=1,3$ (at $m=1,3$ the boundary mode is {\it{exactly}} localized on the outermost row), and
diverges upon approaching the gapless points of the spectrum, i.e., as one approaches $m=0,4$ (towards the trivial phases) or $m=2$, where the Chern number is changed.

\subsection{Correlation matrix, the entanglement spectrum, and the entanglement Hamiltonian}

For a system of free fermions, the entanglement spectrum of a subsystem can be obtained from the
hermitian correlation matrix \cite{Peschel}
\begin{eqnarray}
C_{nm}=\langle c^\dagger_n c^{\phantom{\dagger}}_m \rangle =\Tr \left(\hat{\rho} c_n^\dagger c_m^{\phantom{\dagger}} \right)\;,
\end{eqnarray}
where $c_n$ are fermionic operators, and the site/orbital labels $m,n$ are restricted to be within
the subsystem.  Here $\hat{\rho}$ is the reduced density matrix of the subsystem.  For
non-interacting fermionic systems, all higher order correlation functions can be expressed in terms
of the one-particle correlation matrix.  Following common practice \cite{Alexandradinata@al,
  HughesProdanBernevig, FangGilbertBernevig_PRB2013, TurnerZhangVishwanath_PRB2010,
  ProdanHughesBernevig_PRL2010, HuangArovas_PRB2012, KargarianFiete_PRB2010}, we show the spectra
$\{\xi_j\}$ of the correlation matrix $C$ and even refer to it as the entanglement spectrum.  This
is a `single-particle' version of the entanglement spectrum that is studied in interacting systems.

The reduced density matrix is related to the entanglement Hamiltonian $\hat{\mathcal{H}}$ via 
\begin{eqnarray}
\hat{\rho} = \kappa \exp \left(-\hat{\mathcal{H}} \right)
\end{eqnarray}
where $\kappa$ is a normalization constant.  For free fermions, the entanglement Hamiltonian also
takes a quadratic form
\begin{eqnarray}
\hat{\mathcal{H}} = \sum_{nm} h_{nm} c^\dagger_n c^{\phantom{\dagger}}_m  \;.
\end{eqnarray}
It can be shown \cite{Peschel} that the eigenvalues $\xi_i$ of the correlation matrix $C$ and the
eigenvalue $\epsilon_i$ of the entanglement Hamiltonian $\hat{\mathcal{H}}$ are in one-to-one
correspondence via
\begin{eqnarray}\label{eq: Peschel relation}
\xi_i=\frac{1}{e^{\epsilon_i}+1}\;.
\end{eqnarray}
The correlation matrix, if thought of as an operator, is also of quadratic form:
\begin{align}
\label{eq:corr mat}
\hat C =\sum_{n,m}C_{n,m} c_n^\dagger c_m  = \sum_j \xi_j \gamma^\dagger_j \gamma_j
\end{align}
with $\gamma^\dagger_n=\sum_{j}\bar u^n_j c_j^\dagger$ obtained through a unitary transformation
that diagonalizes the correlation matrix.  One then finds 
\begin{align}
\label{eq:ent ham}
\hat{\mathcal{H}} &=\sum_j \log(\xi^{-1}_j-1)\gamma^\dagger_j \gamma_j \nonumber\\
&=\sum_{n,m}\left(\sum_j \log(\xi^{-1}_j -1) \bar u^j _n u^j_m \right) c^\dagger_n c_m \nonumber\\ 
\end{align}
so that $h_{nm}= \sum_j \log(\xi^{-1}_j -1) \bar u^j _n u^j_m $.  Clearly, the operator
$\hat{C}$ can be interpreted as a hopping Hamiltonian with properties that are qualitatively
similar to the entanglement Hamiltonian.

\section{Entanglement spectrum of a Chern insulator} \label{sec:ES}

\subsection{Entanglement spectrum of a Chern insulator on a torus}

\begin{figure}[t]
\includegraphics[width=0.49\textwidth]{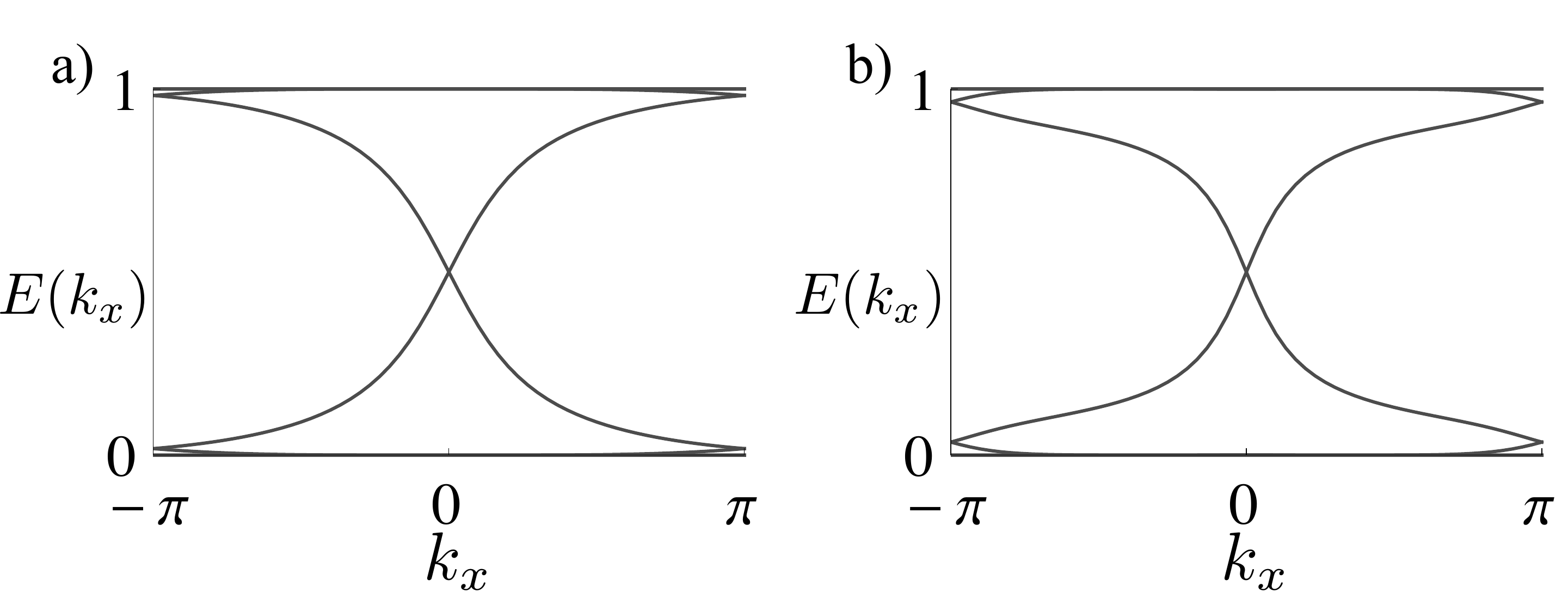}
\caption{\label{fig:estorus}
  Entanglement spectrum of a Chern insulator with a) $m=1$ and b) $m=1.5$ on a torus geometry. The
  occurrence of two boundary modes is due to the existence of two boundaries.
}
\end{figure}

We start with a short reminder of the entanglement spectrum for a cut on a torus geometry. On the
torus there are two boundaries (cuts) between the two subsystems. One immediate consequence of this is the
existence of two boundary modes (one chiral mode living at each boundary).  
Two typical entanglement spectra for a cut on the
torus for $m=1$ and $m=1.5$ are shown in Fig.~\ref{fig:estorus}. Note that there are many degenerate modes at
energies 0 and 1, while the gap-crossing chiral modes are non-degenerate except at the crossing point
at energy 1/2. This crossing point is protected by inversion symmetry as long as the system is cut into two
partitions of equal size \cite{HughesProdanBernevig}. When reducing one of the two partitions while
keeping the total system size fixed, the two virtual edge modes generically hybridize and open a
gap, except at the special points $m=1,3$, where the chiral edge modes persist even when one of
the partitions becomes arbitrarily small.

\subsection{Entanglement spectrum of a Chern insulator in a cylinder geometry}\label{sec:escylinder}

We start with stating the main expectations for the entanglement spectrum of the system in
Fig.~\ref{fig:cylinder} with open boundary conditions. Here, we focus solely on the properties in the topological phase:

(i) Instead of observing two edge modes connecting the upper and lower bands as in the torus
geometry, compare Fig.~\ref{fig:estorus}, we expect one edge mode if the cut is performed far away
from the physical boundary since then we only probe the bulk of the system\cite{Alexandradinata@al}.

(ii) Upon the cut approaching the physical edge the entanglement spectrum should detect that the
system with open boundaries is not a topological insulator but instead a trivial insulator and
consequently a gap should open in the entanglement spectrum, as expected for a trivial insulator.

(iii) The opening of the gap in the entanglement spectrum should be sensitive to the localization
length of the boundary mode.

\begin{figure*}[t]
\includegraphics[width=0.97\textwidth]{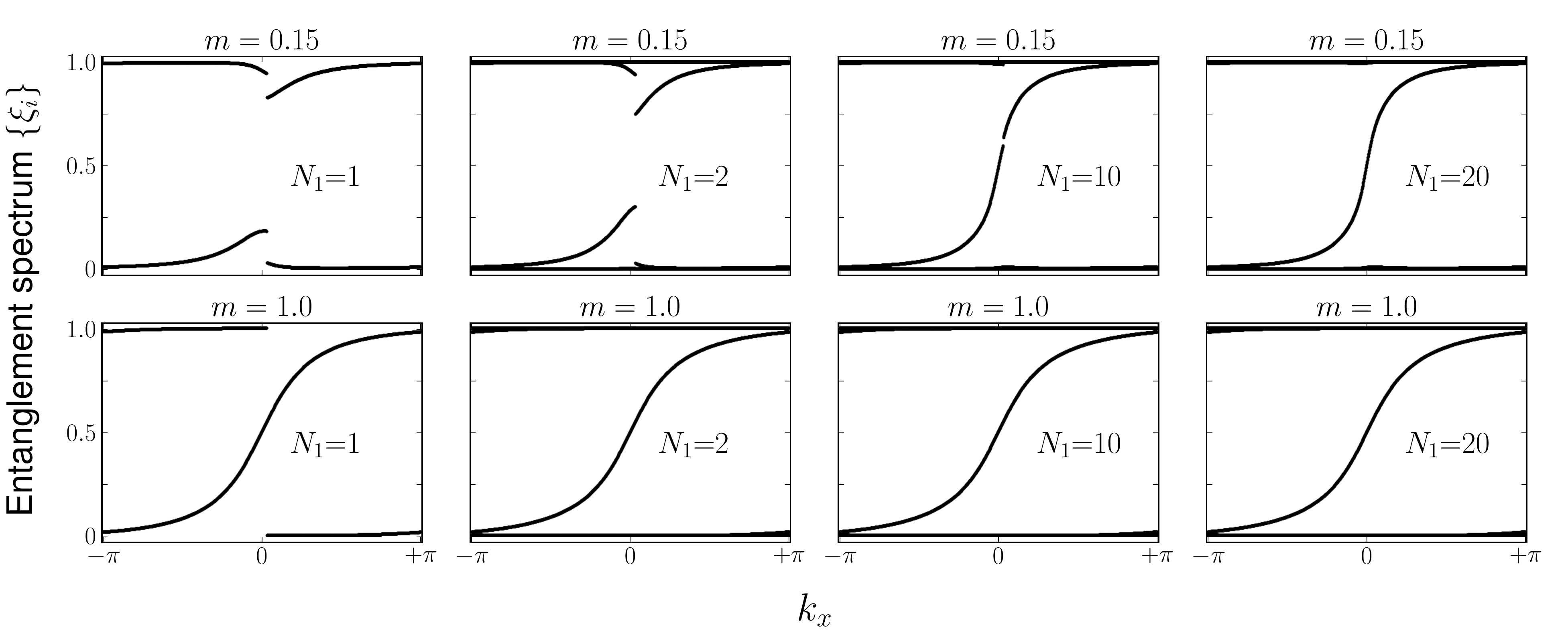}
\caption{\label{fig:mdep} 
Entanglement spectrum for a system of lateral dimension $N=40$, i.e., 40 rows parallel to the
edges. The mass parameter is chosen as $m=0.15$ (a `generic' value) in the top four panels, and as
$m=1$ in the bottom four panels.  We  show the effect of moving the cut away from the boundary,
i.e., increasing $N_1$.}
\end{figure*}

In order to support the first two points we have studied the entanglement spectrum upon varying the
mass parameter $m$ as well as the distance towards the physical boundary. The generic behavior,
exemplified by choosing the mass parameter $m=0.15$ is shown in the top row of Fig.~\ref{fig:mdep}.
We observe that there seems to be one gapless mode connecting the lower and upper bands when
the cut is far away from the physical edge, i.e., when the number of rows $N_1$ in the smaller
partition is significantly larger than 1.  Significantly, a gap opens in the entanglement spectrum
upon approaching the physical edge.

The qualitative behavior is different for the special point $m=1,3$ where the edge mode is
spatially maximally restricted.  This is shown in the second row of Fig.~\ref{fig:mdep}.  The
aforementioned gap in the entanglement spectrum does not open, even for $N_1=1$, i.e., even if the
cut is performed in the row adjacent to the system edge where the physical edge state is
hosted. This happens since for this special value of the mass parameter the edge state is not
localized \emph{exponentially} at the physical edge, but instead \emph{exactly}, which can be traced
back to the flat dispersion of the lower edge of the bulk spectrum, shown in
Fig.~\ref{fig:spectrumob}. 

We note that the Hamiltonian corresponding to the entanglement spectrum seemingly supports a single
chiral channel realized in a one dimensional model.  Since chiral modes are generally thought to
occur in pairs in physical Hamiltonians, this is somewhat peculiar.  We will investigate this aspect
further in Sec.~\ref{sec:eham}.

\begin{figure}[t]
\includegraphics[width=0.4\textwidth]{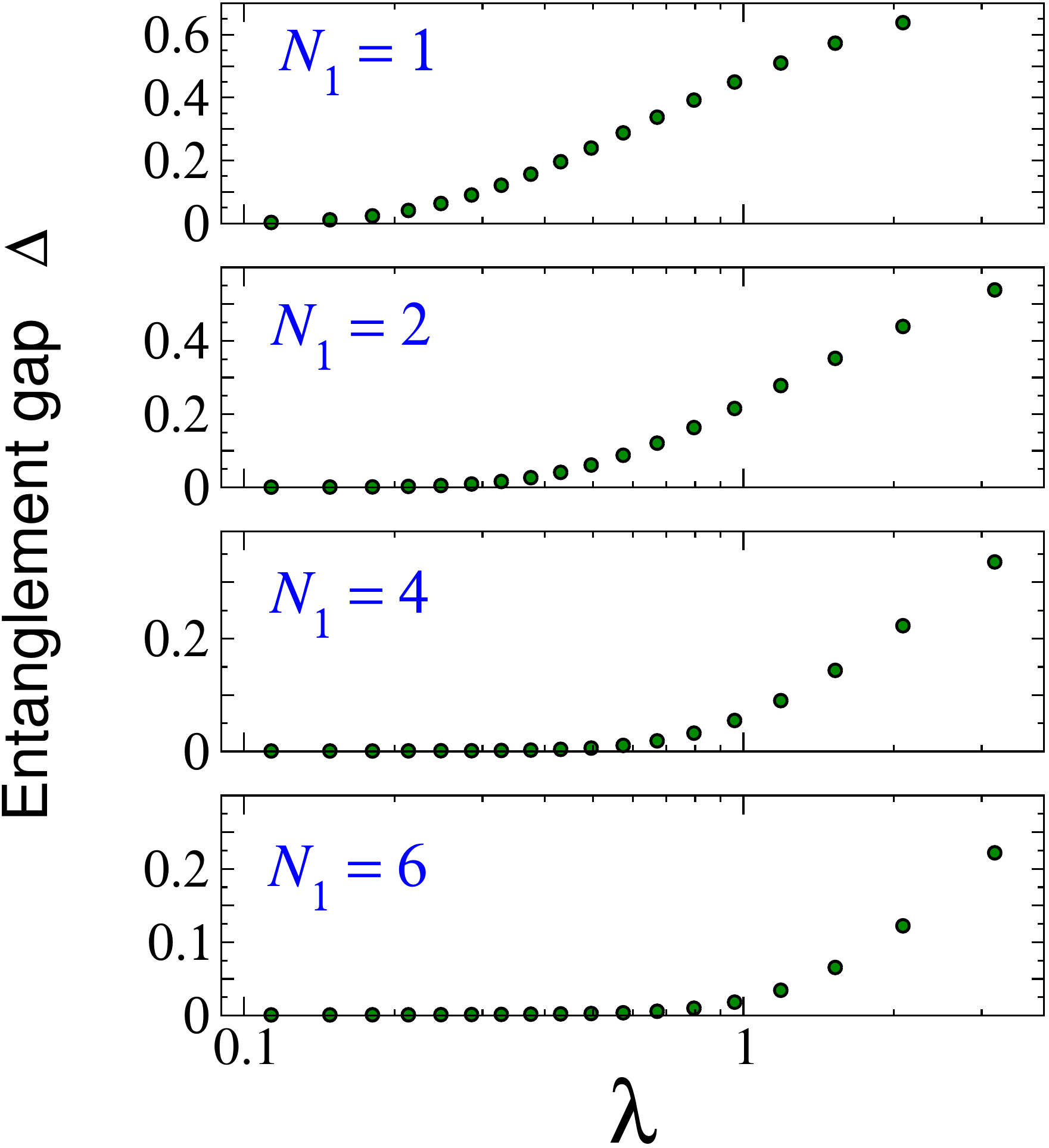}
\caption{\label{fig:gap_versus_loclength} Gap as function of localization length  $\lambda$.  Each point is
  obtained by calculating  $\lambda$ and the entanglement spectrum for a different $m$ value.}
\end{figure}

The foregoing discussion showed the sensitivity of the gap in the entanglement spectrum to the
spatial extent of the physical edge state and the relative position of the cut.  We will now
demonstrate this more quantitatively.  The localization length $\lambda$ can be tuned by varying the
mass parameter $m$, as discussed previously (Figure \ref{fig:loclength_vs_m}).
Figure~\ref{fig:gap_versus_loclength} shows the entanglement gap as a function of $\lambda$, for
partitions containing one, two, four, and six rows.

In the dependence of the entanglement gap on the boundary localization length, we observe
`activated' behavior, roughly when $\lambda$ reaches one-eighth of the number of rows in the
partition.  In other words, the entanglement gap is almost zero when the partition contains almost
all the weight of the boundary mode, but becomes significant when a significant amount of the weight
is outside the partition.

The `activation' part of these curves are reasonably well described by a function of the form
$\Delta(\lambda) \sim \exp\left[\frac{-A_0}{A_1^\epsilon+\lambda^\epsilon}\right]$, with the
exponent $\epsilon \approx 3$.  The length scale of activation defined in this way, $A_1$, is found
to be approximately one-sixth of the number of rows in the partition.

\section{\label{sec:eham}Single-row entanglement Hamiltonian}  \label{sec:entHam}
Let us now discuss some general properties of the correlation matrix as defined in \eqref{eq:corr
  mat}. We consider the case where the system is cut after a single site and is effectively
one-dimensional. The reduced correlation matrix is a hermitian matrix and thus can be used to define
a one-dimensional hopping Hamiltonian $\hat C$.  In the following this Hamiltonian will be denoted as the
entanglement Hamiltonian.  This is a minor abuse of notation as the entanglement Hamiltonian $\hat
{\mathcal{H}}$ is defined by \eqref{eq:ent ham} instead.  However, regarding the correlation matrix
$\hat C$ as the entanglement Hamiltonian is consistent with the common practice \cite{Fidkowski,
  Alexandradinata@al, HughesProdanBernevig} of focusing on $\{\xi_j\}$, which are the eigenvalues of
$\hat C$, instead of on the entanglement spectrum itself. Note that we can always reconstruct the
full many-body entanglement spectrum from the eigenvalues of $\hat C$, implying that all the
information about the system is already captured in the correlation matrix.

As the system is translationally invariant, we can denote the hopping amplitudes of $\hat C$ by
$C_{\alpha\beta}(r)=\phi_{\alpha\beta}(r)|C_{\alpha\beta}(r)|$, where $\alpha$ and $\beta$ denote
the orbital degree of freedom and $r$ is the distance between the sites (in units of the lattice
constant).

\begin{figure}[t]
\includegraphics[width=0.48\textwidth]{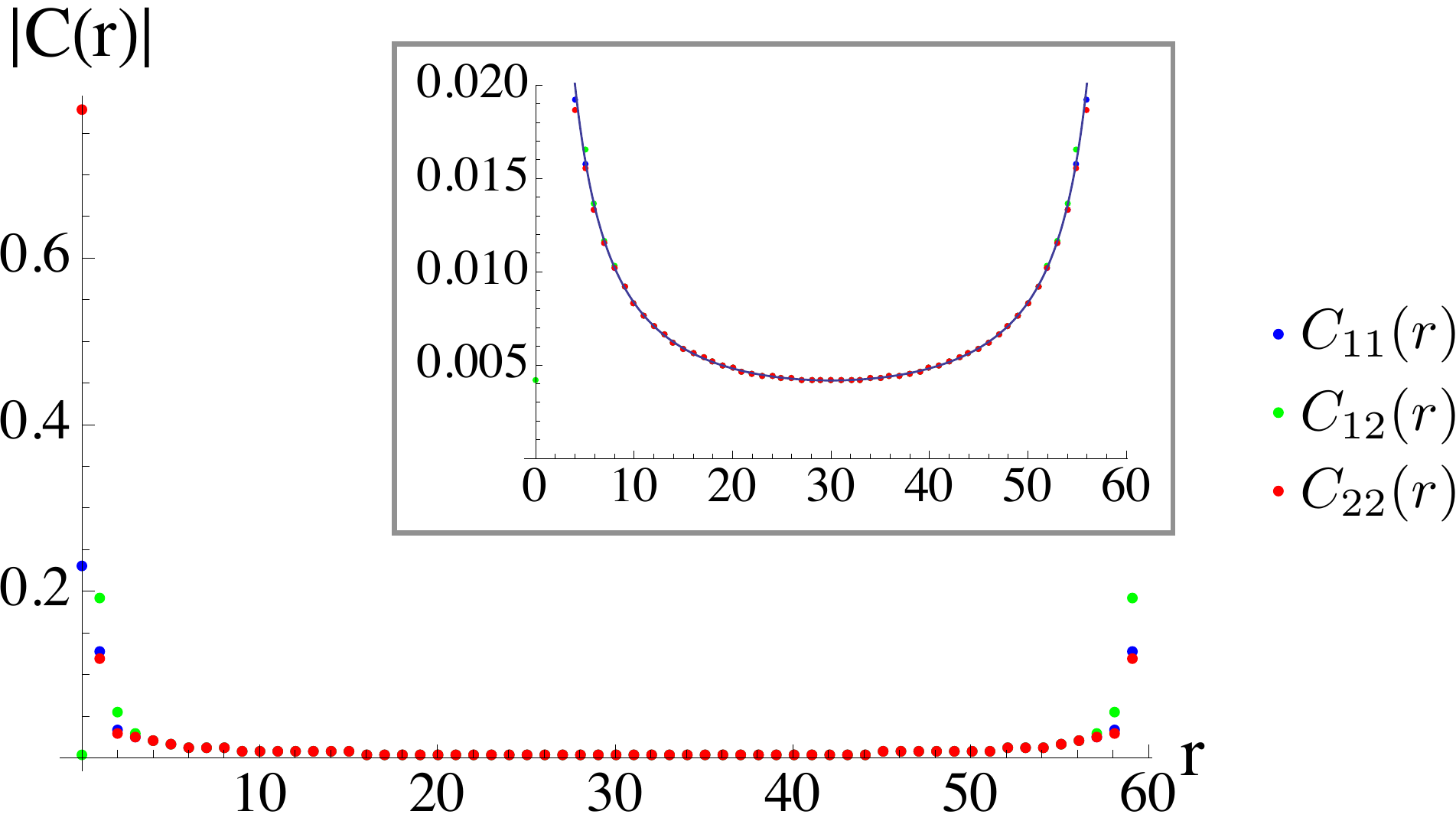}
\caption{ \label{fig:amplitude} Absolute value of the hopping amplitudes in the entanglement
  Hamiltonian \eqref{eq:corr mat} as a function of distance $r$. The solid line in the inset is a fit
  $f(r)=\frac{1}{4N_xSin(\frac \pi{N_x}r)}$ for mass $m=1$.}
\end{figure}

We find that, everywhere in the topological regime, the entanglement Hamiltonian is long-ranged,
even though the original Hamiltonian contains only nearest-neighbor hopping.  For large systems the
magnitudes of the hopping coefficients fall off as the inverse of the distance between the two
sites.  The absolute values of the hopping amplitudes for large distances are well described by
\begin{align}
\label{eq:AbsAmplitude}
|C_{\alpha\beta}(r)|\approx \frac {a(m)} {4N_x \sin(\frac \pi {N_x}|r|)} \;  \mbox{, for } 1\ll r \ll N_x 
\end{align} 
as shown in Figure \ref{fig:amplitude}. When the system \eqref{eq:model} is in the trivial regime, however, the corresponding entanglement Hamiltonian is always short-ranged. This behavior is not unexpected, as the ground state in the trivial  phase can always be adiabatically connected to the atomic insulator, which by definition has a short-range entanglement Hamiltonian (see also Ref. \cite{long-range}).

\begin{figure}[t]
\includegraphics[width=0.48\textwidth]{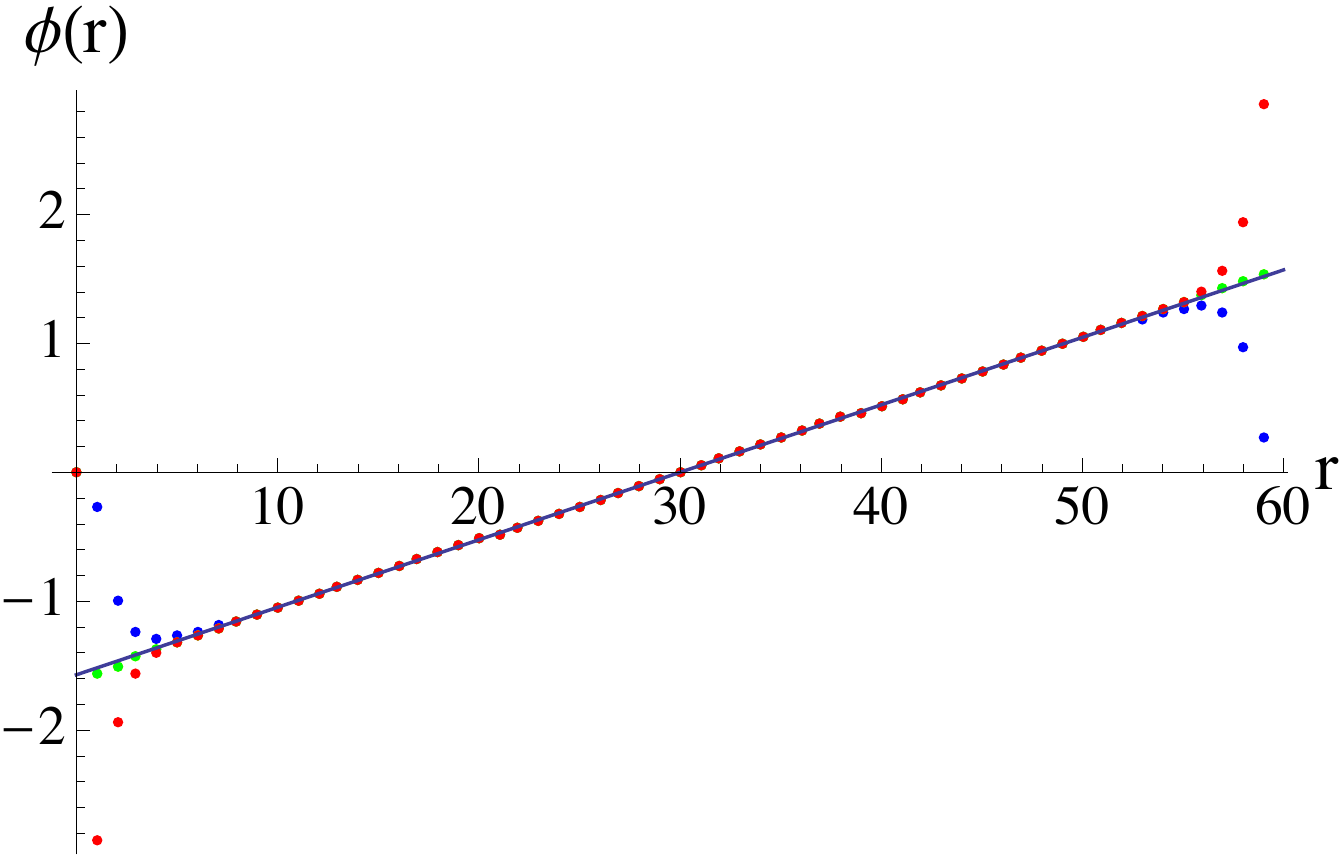}
\caption{(Color online) Phase value of the hopping amplitudes in the entanglement Hamiltonian as a
  function of distance $r$. The solid line is a fit $f(r)=\frac{\pi}{N_x}r-\frac \pi 2$ for mass
  $m=1$. \label{fig:phase}
}
\end{figure}

Let us also note the remarkable fact that the phases $\phi_{\alpha\beta}(r)$ of the long-range
hopping amplitudes are highly correlated for large values of $r$ and independent of the orbital
indices. More specifically, away from the phase transition points $\phi_{\alpha\beta}(r)$ can be very well approximated by a linear function of $r$:
\begin{align}
\label{eq:PhaseAmplitude}
\phi_{\alpha\beta}(r)\approx \frac{2-m}{|2-m|} \left(\frac \pi  {N_x} r-\frac \pi 2\right)\mbox{, for }1\ll r \ll N_x.
\end{align} 
An example of this behavior is seen in Fig.~\ref{fig:phase}, where $\phi_{\alpha\beta}(r)$ is
plotted for $m=1$ and system size $N_x =N_y=60$.  There is some deviation at small distances from
the exact linearity.

An interesting property of this Hamiltonian is that it shows the phenomenon of a so-called persistent
current \cite{persistentcurrent}. A simple way to see this is from the spectrum itself, shown in Fig.~\ref{fig:mdep} (second
row, leftmost panel). The current can be determined as a summation of all the velocities $j \propto
dE/dk$ for occupied states. Irrespective of the chemical potential we will only sum positive group
velocities meaning there will always be a finite net current, leading to the persistent
current.

We have seen in Sec.~\ref{sec:escylinder} that for $m=1$ the enanglement spectrum is not gapped even
if the cut is performed in the row adjacent to the physical edge mode. Consequently, the spectrum
looks as if it was a one dimensional chiral Hamiltonian. This seems contradictory to the fermion
doubling theorem for systems with chiral and translational symmetry~\cite{NielsenNinomiya1981}, but
the way out is via long-range hoppings ($\sim \frac 1 r $) which effectively mimics a higher
dimensional system.

\section{Flux insertion into the cylinder}\label{sec:flux}

Let us now discuss what happens when magnetic flux is threaded through the cylinder. This can be
done in several, inequivalent ways. Following Ref. \onlinecite{Alexandradinata@al} the flux can be
inserted in the original Hamiltonian. In this scenario, the Hamiltonian is invariant under $2\pi$
flux insertion, but the correlation matrix (and hence the entanglement Hamiltonian) are not, because
the ground state evolves into an excited state. Under flux insertion the spectrum of the correlation matrix
changes such that effectively one state is transported from $\xi=0$ to $\xi=1$ (or vice versa
depending on the direction of the flux), but all other states transform into each other. 
Despite this change in the correlation matrix (and hence the entanglement spectrum) 
the authors of Ref. \onlinecite{Alexandradinata@al} concluded that the entanglement spectrum shows
spectral flow.

However, when considering the correlation matrix as a hopping Hamiltonian, we can also insert flux
in this effective Hamiltonian. This is not equivalent to the scheme explained above, because in this
case, the spectrum has to be invariant under $2\pi $ flux insertion. In particular, we are
interested how the flux insertion affects the spectrum at the fine-tuned points $m=1,3$, where the
entanglement spectrum appears gapless with a single chiral channel.
The appearance of a single chiral channel is worrisome as it suggests that a level can be
transported upon flux insertion from the lower to the upper band, while the reverse process is not
possible.  This implies that the system cannot go back to itself upon adiabatic insertion of flux
$2\pi$ if the upper and the lower band are indeed connected. Consequently, the existence of a chiral
channel must be deceiving.

This can be seen most clearly when the mass parameter $m\neq 1,3$. When the cut approaches the edge,
i.e. the entanglement Hamiltonian becomes more and more one-dimensional, a gap opens in the
entanglement spectrum. In this case, adiabatic flux insertion is expected to act separately on the
upper and lower band and there is no spectral flow in the entanglement spectrum.  At the special
points the situation is more intriguing since there is no gap in the entanglement spectrum and a more 
thorough analysis of the effects of flux insertion is needed in order to determine
whether or not adiabatic level transport between the bands is possible. 

A consistent way to introduce flux in expression \eqref{eq:corr mat} and \eqref{eq:ent ham} is by
changing the boundary conditions from periodic to $c_{\alpha,N_x}^{\dagger}\equiv e^{i\phi}
c_{\alpha,0}^\dagger$.  The correlation matrix is written as a sum over the hopping range $r$:  
\begin{align}
\hat C &=\sum_{\alpha,\beta}\sum_{j=0}^{N_x-1} \sum_{r=0}^{\lfloor (N_x -1)/2\rfloor} C_{\alpha,\beta}(r) c^\dagger_{\alpha,j+r}c_{\beta,j}+h.c.\nonumber\\
&+\frac {1+(-1)^{N_x}} 4  \sum_{\alpha,\beta}\sum_{j=0}^{N_x-1}  C_{\alpha,\beta}\left(\frac {N_x}{2}\right) c^\dagger_{\alpha,j+N_x/2}c_{\beta,j}+h.c.
\end{align}
and the boundary condition is consequently applied every time $j+r\geq N_x$.  The factor $1/4$ in
the second line is needed to avoid double-counting.  In this setup, the entanglement spectrum never
shows spectral flow for finite size systems. When the Hamiltonian is not tuned to $m=1,3$, there is
a gap in the spectrum and the eigenstates of the lower and upper band transform into each other
separately, as expected. In the fine-tuned case, the situation is more complicated. However, for any
finite system sizes, there is a finite-size gap preventing the transport of a level from the lower
to the upper band. In the thermodynamic limit, the entanglement spectrum becomes discontinuous due
to the long-range hopping and the special phase relations shown in Figs. \ref{fig:amplitude} and
\ref{fig:phase}. In this case, the chiral mode transports a state from the lower to the upper band,
but at the same time another state is transported from the upper to the lower band via the band
discontinuity. As a result, the entanglement spectrum shows spectral flow in the thermodynamic limit
at the special points $m=1,3$.  As a final comment we should note, that inserting flux in the
correlation matrix and the entanglement Hamiltonian is not equivalent in this setup. In particular,
the eigenvalues are no longer connected by Eq. \eqref{eq: Peschel relation}, when the flux is
non-zero.

\section{Conclusion}   \label{sec:concl}

In this work we have investigated the entanglement spectrum of Chern insulators in a half-open
geometry as a function of the distance of the entanglement cut to the physical edge of the system.
In contrast to the by-now standard entanglement studies which create and study \emph{virtual} edge
modes at the entanglement cut, this unusual setup is designed to study effects of the
\emph{physical} edge mode at the physical boundary on the entanglement spectrum and the entanglement
Hamiltonian.  In the extreme case where one of the partitions contains only a single row, the
virtual and physical edges are the same.  This is reminiscent of entanglement spectra studies of
(non-topological) ladder systems with the cut between the two ladder legs \cite{ladder}.
We presented consequences of the interplay between the spatial extent of the edge mode and the cut
position, exploiting the tunability of the model Hamiltonian to also examine special points where
the physical edge mode is sharply localized.

Strictly speaking, in such a geometry the system is either a trivial insulator, if the edge modes
are exponentially localized, or a metal if the edge modes are localized exactly.  In the first
case we observe an opening of the gap in the entanglement spectrum upon approaching the physical
edge, while in the second case we find a seemingly single chiral channel.  In the former case we
found that no level transport across the gap is permitted accounting for the fact that the system
with open boundaries is topologically trivial. In the latter, level transport is prohibited for any
finite size system.   In the thermodynamic limit, level transport between the upper
and lower band is allowed at the expense of the spectrum becoming discontinuous.
 
We also analyzed the one dimensional entanglement Hamiltonian, obtained by placing the cut directly
at the edge, in more detail. The resulting Hamiltonian, living on a closed chain, has remarkable
properties, such as long-range hopping amplitudes with highly correlated complex phase factors. In
addition, the ground state of this Hamiltonian hosts a persistent current.  While the details depend
on the value of the mass parameter, the generic properties remain the same throughout the
topological phase.  This persistent current in the entanglement Hamiltonian ground state is a
reflection the virtual chiral edge mode of the subsystem.
For generic mass parameters, the effective one-dimensional system is gapped.  The corresponding
spectrum is that of an insulator if the chemical potential is in this gap, although the notion of
chemical potential for an entanglement Hamiltonian is somewhat ambiguous.

The question remains if these properties are a peculiarity of the particular model we chose or if
they may be more general. In order to investigate this, we studied the properties of another simple
topological insulator model on the checkerboard lattice, first introduced in
Ref. \cite{checkerboard}. Even though the details are different, we find that the main features
remain the same. In particular, we find that the gapless, seemingly chiral channels can again
persist even when cutting the system directly at the physical edge. The resulting one-dimensional
entanglement Hamiltonian is long-range.  The dependence of the phases of the complex hopping
amplitudes on hopping distance is more complicated than the present case, but the overall trend is
similar.  This system also seems to support a persistent current.

Given this supporting information from a second topological insulator model, we believe that the
features we have presented in this work are generic to Chern insulators.  An intriguing conjecture
is that analogous features might also be generic to \emph{interacting} topological states.  It is
also interesting to ask how these results for 2D topological systems are generalized and modified
for different classes of 3D topological insulators.


\begin{acknowledgments}
  
  The authors acknowledge useful discussions with E.~Ardonne, E.~Bergholtz and F.~Pollmann.  This
  work was supported by the "Deutsche Forschungsgemeinschaft" within unit SFB TR12 (M.~Hermanns) and
  the Emmy-Noether program FR 2627/3-1 (LF).  This work is also part of the D-ITP consortium, a
  program of the Netherlands Organisation for Scientific Research (NWO) that is funded by the Dutch
  Ministry of Education, Culture and Science (OCW)
\end{acknowledgments}


\begin{thebibliography}{33}

\bibitem{Classification1}
S.~Ryu and A.~P.~Schnyder and A.~Furusaki and A.~W.~W.~Ludwig, New J.~Phys. {\bf 12}, 065010 (2010). 

\bibitem{Classification2}
A.~Kitaev, AIP Conf. Proc. \textbf{1134}, 22 (2009).


\bibitem{IQH}
K.~v.~Klitzing, G.~Dorda, and M.~Pepper, Phys. Rev. Lett. {\bf 45}, 494 (1980). 


\bibitem{Haldane1988}{%
F.~D.~M.~Haldane, Phys.\ Rev.\ Lett.\ \textbf{61}, 2015 (1988).
}

\bibitem{TopIns1}
C.~L.~Kane and E.~J.~Mele, Phys. Rev. Lett. {\bf 95}, 226801 (2005); Phys. Rev. Lett. 95, 146802 (2005). 

\bibitem{TopIns2}
B.~A.~Bernevig, T.~L.~Hughes, and S.-C.~Zhang, Science {\bf 314}, 1757 (2006). 

\bibitem{TopIns3}
R.~Roy, Phys. Rev. B {\bf 79}, 195321 (2009). 


\bibitem{Exp1}
M.~K{\"o}nig, S. Wiedmann, C.~Br{\"u}ne, A.~Roth, H.~Buhmann, L.~W.~Molenkamp, X.-L.~Qi,
S.-C.~Zhang, Science {\bf 318}, 766 (2007).  

\bibitem{Exp2}
D.~Hsieh,  D.~Qian, L.~Wray, Y.~Xia, Y.~S.~Hor, R.~J.~Cava, and M.~Z.~Hasan, Nature {\bf 452}, 970 (2008). 

\bibitem{LiHaldane}
H.~Li  and F.~D.~M.~Haldane, Phys. Rev. Lett. {\bf 101}, 010504 (2008). 

\bibitem{ES_in_FQH} 
%
O.~S.~Zozulya, M.~Haque, and N.~Regnault, Phys. Rev. B \textbf{79}, 045409 (2009). 
%
A.~M.~L\"auchli, E.~J.~Bergholtz, J.~Suorsa, and M.~Haque, Phys.\ Rev.\ Lett.\ \textbf{104}, 156404 (2010). 
%
R.~Thomale, A.~Sterdyniak, N.~Regnault, and B.~A.~Bernevig, Phys. Rev. Lett. \textbf{104}, 180502 (2010).
%
R.~Thomale, D.~P.~Arovas, and B.~A.~Bernevig, Phys. Rev. Lett. \textbf{105}, 116805 (2010).
%
A.~Sterdyniak, B.~A.~Bernevig, N.~Regnault, and F.~D.~M.~Haldane, New J.~Phys. \textbf{13}, 105001 (2011).  
%
Z.~Papic, B.~A.~Bernevig, and N.~Regnault, Phys. Rev. Lett. \textbf{106}, 056801 (2011). 
%
A.~Sterdyniak, N.~Regnault, and B.~A.~Bernevig, Phys. Rev. Lett. \textbf{106}, 100405 (2011).
%
A.~Chandran, M.~Hermanns, N.~Regnault, and B.~A.~Bernevig,  Phys. Rev. B \textbf{84}, 205136 (2011).   
%
M. Hermanns, A. Chandran, N. Regnault, and B. Andrei Bernevig, Phys. Rev. B \textbf{84}, 121309(R)
%
J. Biddle, M.~R.~Peterson, and S.~Das~Sarma, Phys. Rev. B \textbf{84}, 125141 (2011).  
%
 J. Zhao, D. N. Sheng, and F. D. M. Haldane, Phys. Rev. B \textbf{83}, 195135 (2011). 
%
J.~Schliemann, Phys. Rev. B \textbf{83}, 115322 (2011).
%
A.~Sterdyniak, B.~A.~Bernevig, N.~Regnault, and F.~D.~M.~Haldane, New J.~Phys. \textbf{13}, 105001 (2011). 
%
X.-L.~Qi, H.~Katsura, and A.~W.~W.~Ludwig, Phys. Rev. Lett. \textbf{108}, 196402 (2012).
%
I.~D.~Rodriguez, S.~H.~Simon, and J.~K.~Slingerland, Phys. Rev. Lett. \textbf{108}, 256806 (2012).  
%
Z.~Liu, E.~J.~Bergholtz, H.~Fan, and A.~M.~L\"auchli, Phys. Rev. B \textbf{85}, 045119 (2012). 
%
A.~Sterdyniak, A.~Chandran, N.~Regnault, B.~A.~Bernevig, and P.~Bonderson, Phys. Rev. B \textbf{85}, 125308
(2012). 
%
A.~Sterdyniak, N.~Regnault, and G.~M\"oller, Phys. Rev. B \textbf{86}, 165314 (2012).
%
M.~P.~Zaletel and R.~S.~K.~Mong, Phys. Rev. B \textbf{86}, 245305 (2012).
%
J.~Dubail, N.~Read, and E.~H.~Rezayi, Phys. Rev. B \textbf{86}, 245310 (2012).  
%
Y.~Wu, N.~Regnault, and B.~A.~Bernevig, Phys. Rev. Lett. \textbf{110}, 106802 (2013).
%
M.~P.~Zaletel, R.~S.~K.~Mong, and F.~Pollmann,  Phys. Rev. Lett. \textbf{110}, 236801 (2013).  
%
S.~Furukawa and M.~Ueda, Phys. Rev. Lett. \textbf{111}, 090401 (2013). 
%
Z.~Liu and E.~J.~Bergholtz,  Phys. Rev. B  \textbf{87}, 035306 (2013). 
%
I.~D.~Rodriguez, S.~C.~Davenport, S.~H.~Simon, and J.~K.~Slingerland, Phys. Rev. B \textbf{88}, 155307
(2013).

 

\bibitem{ES_in_interacting_nonFQH}
%
H.~Yao and X.-L.~Qi, Phys. Rev. Lett. \textbf{105}, 080501 (2010). 
%
F.~Pollmann, A.~M.~Turner, E.~Berg, and M.~Oshikawa, Phys. Rev. B \textbf{81}, 064439 (2010).
%
J.~I.~Cirac, D.~Poilblanc, N.~Schuch, and F.~Verstraete, Phys.\ Rev.\ B \textbf{83}, 245134 (2011). 
%
N.~Regnault and B.~A.~Bernevig, Phys. Rev. X {\bf 1}, 021014 (2011). 
%
T.~Grover, arXiv:1112.2215. 
%
Y.-L.~Wu, B.~A.~Bernevig, and N.~Regnault, Phys. Rev. B \textbf{85}, 075116 (2012). 
%
B.~A.~Bernevig and N.~Regnault, Phys. Rev. B \textbf{85}, 075128 (2012). 
%
D.~Poilblanc, N.~Schuch, D.~P{\'e}rez-Garc{\'i}a, and J.~I.~Cirac,   Phys. Rev. B \textbf{86}, 014404 (2012).
%
D.~Poilblanc and N.~Schuch, Phys. Rev. B \textbf{87}, 140407(R) (2013).
%
T.~Liu, C.~Repellin, B.A.~Bernevig, and N.~Regnault, Phys. Rev. B \textbf{87}, 205136 (2013). 
%
A.~Sterdyniak, C.~Repellin, B.~A.~Bernevig, and N.~Regnault, Phys. Rev. B \textbf{87}, 205137 (2013). 
%
W.~Zhu, D.~N.~Sheng, and F.~D.~M.~Haldane, Phys. Rev. B \textbf{88}, 035122 (2013).  
%
Z.~Liu, D.~L.~Kovrizhin, and E.~J.~Bergholtz, Phys. Rev. B \textbf{88}, 081106(R) (2013). 
%
Z.~Liu, E.~J.~Bergholtz, and E.~Kapit, Phys. Rev. B \textbf{88}, 205101 (2013).

\bibitem{Fidkowski}
L. Fidkowski, Phys. Rev. Lett. {\bf 104}, 130502 (2010). 

\bibitem{EntSpecIntTI1}
A.~M.~Turner, Y.~Zhang, and A.~Vishwanath, Phys. Rev. B {\bf 82}, 241102(R) (2010).

\bibitem{HughesProdanBernevig}{
T.~L.~Hughes, E.~Prodan, and B.~A.~Bernevig, Phys. Rev. B {\bf 83}, 245132 (2011).
}

\bibitem{Alexandradinata@al}
{
A.~Alexandradinata, T.~L.~Hughes, and B.~A.~Bernevig, Phys. Rev. B {\bf 84}, 195103 (2011).
}

\bibitem{FangGilbertBernevig_PRB2013} C.~Fang, M.~J.~Gilbert, and B.~A.~Bernevig, Phys.\ Rev.\ B
  \textbf{87}, 035119 (2013).

\bibitem{TurnerZhangVishwanath_PRB2010} A.~M.~Turner, Y.~Zhang, and A.~Vishwanath, Phys.\ Rev.\ B
  \textbf{82}, 241102(R) (2010).


\bibitem{ProdanHughesBernevig_PRL2010} E.~Prodan, T.~L.~Hughes, and B.~A.~Bernevig, Phys.\ Rev.\
  Lett.\ \textbf{105}, 115501 (2010).

\bibitem{KargarianFiete_PRB2010} M.~Kargarian and G.~A.~Fiete, Phys.\ Rev.\ B \textbf{82}, 085106
  (2010).

\bibitem{HuangArovas_PRB2012} Z.~Huang and D.~P.~Arovas, Phys.\ Rev.\ B \textbf{86}, 245109 (2012).

\bibitem{persistentcurrent}
M.~B{\"u}ttiker, Y.~Imry and R.~Landauer, Phys. Lett. A \textbf{96}, 365 (1983). 

\bibitem{spectralflow}
A.~M.~Turner, Y.~Zhang, A.~Vishwanath, Phys. Rev. B \textbf{82}, 241102R (2010).


\bibitem{Peschel}{
I.~Peschel, J.~Phys.~A: Math.\ Gen. \textbf{36}, L205 (2003). 
}

\bibitem{long-range}
I. Peschel and M.-C. Chung,  EPL \textbf{96}, 50006 (2011).

\bibitem{NielsenNinomiya1981} H.~B.~Nielsen and M.~Ninomiya, Phys.\ Lett.\ B {\bf{105}}, 219 (1981).



\bibitem{ladder}  D.~Poilblanc , Phys.\ Rev.\ Lett.\ \textbf{105}, 077202 (2010).
%
J.~Schliemann and A.~M.~L{\"a}uchli, J.~Stat.\ Mech.\ P11021 (2012). 
%
R.~Lundgren, Y.~Fuji, S.~Furukawa, and M.~Oshikawa, Phys. Rev. B \textbf{88}, 245137 (2013).

\bibitem{checkerboard} 
R.~Roy,  arXiv:cond-mat/0603271 (2006).



\end{thebibliography}
\end{document}